\documentclass[12pt,preprint]{aastex} 
\slugcomment{} 

\newcommand\kms{\ifmmode {\rm km\ s}^{-1} \else km s$^{-1}$\fi}
\newcommand\eflux{\ifmmode {\rm ergs\ s}^{-1}\;{\rm cm}^{-2} \else  
	ergs s$^{-1}$ cm$^{-2}$\fi}  
\newcommand\phflux{\ifmmode {\rm photons\ s}^{-1}\;{\rm cm}^{-2} 
	\else  	photons s$^{-1}$ cm$^{-2}$\fi}  
\newcommand\ergsec{\ifmmode {\rm ergs\ s}^{-1} \else  
	ergs s$^{-1}$\fi}
\newcommand\Msun{\ifmmode M_{\odot} \else $M_{\odot}$\fi}

\shorttitle{Case for diskline} 
\shortauthors{Turner et al.\ 2001} 
 
\begin{document}
\title{
Narrow Components within the 
Fe K$\alpha$ Profile of NGC 3516: Evidence for the Importance of 
General Relativistic Effects? }

\author{T.\ J.\ Turner\altaffilmark{1,2}, R.\ F.\ Mushotzky\altaffilmark{2},
	T.\ Yaqoob\altaffilmark{1,3}, 
	I.\ M.\  George\altaffilmark{1,2}, 
	S.\ L.\ Snowden\altaffilmark{2,4}, H.\ Netzer\altaffilmark{5}, 
	S.\ B.\ Kraemer\altaffilmark{6}, K.\ Nandra\altaffilmark{2,4}, 
	D.\ Chelouche\altaffilmark{5}
}

\altaffiltext{1}{Joint Center for Astrophysics, Physics Dept., University of Maryland
	Baltimore County, 1000 Hilltop Circle, Baltimore, MD 21250} 
\altaffiltext{2} {Laboratory for High Energy Astrophysics, Code 662, 
	NASA/GSFC, Greenbelt, MD 20771}
\altaffiltext{3}{Center for Astrophysical Sciences, 
	Department of Physics and Astronomy, The Johns Hopkins 
	University, Baltimore, MD 21218}
\altaffiltext{4}{Universities Space Research Association, Suite 206, 
Forbes Boulevard, Seabrook, MD}
\altaffiltext{5}{School of Physics and Astronomy, Raymond and Beverly
Sackler Faculty of Exact Sciences, Tel-Aviv University, Tel-Aviv 69978, Israel}
\altaffiltext{6}{Catholic University of America, NASA/GSFC, Code 681,
Greenbelt, MD 20771}

\begin{abstract}

We present results from a simultaneous {\it Chandra} HETG and 
{\it XMM-Newton} observation of NGC 3516. 
We find evidence for several narrow components of Fe K$\alpha$ 
along with a broad line. 
We consider the possibility that the lines arise in an  blob of material ejected 
from the nucleus with velocity $\sim$ 0.25c.
We also consider an origin 
in a neutral accretion disk, suffering enhanced illumination at 
35 and 175 ${\rm R_g}$, perhaps due to magnetic reconnection. 
The presence of these narrow features 
indicates there is no Comptonizing region along the line-of-sight to 
the nucleus. This in turn 
is compelling support for the hypothesis 
that  broad Fe K$\alpha$ components are, in general, produced by
strong gravity. 

	\end{abstract}


	\section{Introduction}

Active Galactic Nuclei (AGN) are believed to be powered   by  accretion of 
material onto a supermassive black hole.  Intrinsically narrow emission lines 
from the accretion disk 
are predicted to produce double-horned profiles due to the Doppler 
effect  (unless viewed face-on). 
Near the innermost stable orbit the gas 
velocity approaches the speed of light 
and special relativistic beaming should enhance the 
``blue'' peak on the approaching side 
relative to that on the receding  (``red'') side. 
Gravitational and transverse Doppler effects  
redistribute photons to lower energies and 
introduce asymmetric offsets of the Doppler horns from the rest-energy. The sum of 
contributions over a range of annuli then gives a broad (FWHM $\Delta E/E \sim0.3$) 
asymmetric profile (\citealt{L91, fab2000}). 
The X-ray band-pass contains the strongest such line, Fe K$\alpha$, emitted via 
fluorescence or recombination processes  between  6 -- 7 keV,  depending on the 
ionization-state of the gas. This line is commonly observed  
in AGN (e.g., \citealt{N97}) with both narrow 
\citep{tazza} and  broad components (e.g., \citealt{tanaka95});
the former may arise in cool material close to the optical broad-line-region 
while the latter is thought to originate 
close to the black hole (see \citealt{fab2000} and
references therein). 
However, there is some controversy as to the origin of the broad line component, 
with broadening mechanisms such as Comptonization (\citealt{missut,miskemb}) 
being suggested, along with other possibilities 
(e.g., \citealt{schurch,zjm}).

X-ray detectors  prior to {\it Chandra} had 
only moderate energy resolution around 6 keV  (e.g. $\Delta E/E \sim 0.02$) 
leaving the detailed line shape and origin unclear. 
The high-energy-transmission grating spectrometer 
(HETGS; \citealt{markert94}) of  {\it Chandra} gives a factor $\sim 4$ improvement in 
energy resolution at 6 keV 
compared to  detectors flown previously, allowing isolation of contributions to the 
profile from weak narrow features. The EPIC CCDs on board  {\it XMM-Newton} 
(hereafter referred to as  {\it XMM})  yield high throughput, 
allowing tight constraints on the 
continuum and broad line. Together these satellites offer an unprecedented insight 
into the Fe K$\alpha$ profile.  
Here we present  results from 
overlapping {\it Chandra} and {\it XMM}  observations of NGC~3516, 
revealing new features that provide compelling 
evidence  the line is indeed modified by special and general relativistic effects.  

\section{New {\it Chandra} and {\it XMM} Results}

A {\it Chandra} observation of NGC~3516 between 
2001 November 11 UT 01:00:25 -- 12 UT 02:19:22 
overlapped an {\it XMM} 
observation  covering November 09 UT 23:12:51 -- 11 UT 
10:54:19 and a {\it RXTE} observation 
November 11 UT 10:05:36 - UT 11:10:56 (Figure~1). 
{\it Chandra} data were reprocessed using 
{\tt CALBD v2.6} and {\tt CIAO v2.1}, 
removing bad pixels, columns, period of high background and 
events with detector `grades' {\it not} equal to 0, 2, 3, 4, or 6; 
yielding an  exposure of $\sim$75~ks. 
The EPIC data reduction pipeline
was run with SAS v5.2.0. 
%and calibration files from September 2001. 
EPIC data utilized the thin
filter with PN Prime Small Window mode, 
MOS1 Small Window Free Running mode and MOS2 Large Window mode. 
These data were screened to remove hot and bad pixels 
and periods of high background, yielding an exposure of 80 ks 
(20 ks overlapped {\it Chandra}). 
At this flux level the photon pileup is negligible.
Instrument patterns 0 -- 12 (MOS) and 
0 -- 4  (PN) were selected.  
Spectra were extracted from a cell $\sim 0.94'$ diameter, 
centered on the source. Background spectra were extracted from a nearby 
region for the PN ($< 1$\% of the source count rate) 
and ignored for the MOS detectors (the small window mode
in the MOS leaves no nearby regions for background extractions, but 
background is negligible at this flux level). 
{\it RGS} data will be presented in a later paper. 
The {\it RXTE} observation of 3.5 ks overlapped the 
{\it XMM} and {\it Chandra} data as shown (Figure~1). 

\section{The X-ray Spectrum}

We found NGC~3516 to 
have a flux $F_{2-10\ keV} \sim  1.3 - 1.5 \times 10^{-11} {\rm erg\ cm^{-2}\ s^{-1}}$,
the lowest observed, 
first exhibited during a {\it Chandra} LETG observation 
\citep{netzer02}. 
{\it XMM} data reveal a flat spectrum 
across 2-10 keV, with evidence for 
transmission through ionized gas as previously 
observed in this source (e.g., \citealt{netzer02,kraemer02,kriss96}).  
The {\it XMM} light curves and hardness ratio 
suggested a natural division $\sim 60$ks into the observation. 
The second segment of {\it XMM} data  were fit with 
the entire {\it Chandra} dataset during a time-period denoted T2. 
While not ideal (the {\it Chandra} data extending beyond the end of the 
{\it XMM} observation), this division of the data proves enlightening. 
%For  period T1 the continuum photon index was 
%$\Gamma=1.21\pm0.02$, for T2  $\Gamma=1.16\pm0.05$.  

We examined the ratio of Fe K$\alpha$ photons relative to the local
continuum. 
Figure~2a shows the HETG data alone, for clarity. The positive and 
negative data from the first  
order of the HEG are combined with those of the MEG (up to 6.5 keV). 
Fe K$\alpha$ emission is evident with a strong  peak at 
rest-energy  6.41 keV. The associated Fe K$\beta$ line 
is seen at 7.06 keV consistent with the expected strength $\sim 11$\% 
of the K$\alpha$ flux. 
The Fe K$\alpha$ component has width FWHM $\sim 40$ eV and is constant in 
flux over the observation. During the latter part of the {\it XMM}
observation (T2)  PN data hint at a double line peak but MOS 
data do not confirm it and the HETG data are ambiguous. 
The constancy in flux, the width and  energy indicate  
the bulk of the line at 6.4 keV probably arises in cool material 
located at least several  
light-days from the nucleus rather than the accretion flow, thus we 
do not consider this component or the associated K$\beta$ line further. 

Most importantly, additional and unexpected lines are evident at  5.57  keV and 
6.22 keV. The line widths are consistent with zero, with upper limits 
$FWHM \sim 500$ and 40 eV, respectively. The line at 5.6 keV is weak 
and it is difficult to separate it from the broad red wing, hence the 
extremely loose constraint on width.  This fine structure has never 
before been seen in an AGN and the lines do not correspond to any 
emission lines expected of 
appreciable strength.  These are not statistical fluctuations, 
they show up in the positive 
and negative sides of the dispersed spectra from both parts of the 
HETGS. Overlapping 
{\it XMM} data from T2 (Figure 2b) confirm the existence of a feature at 6.2 keV and show 
other features between 6.8 -- 6.9 keV in the {\it XMM} data (Figures 2b, c). 

Table 1 details line fluxes obtained from the HETGS and two subsets of 
{\it XMM} data,  revealing variability in the lines at 5.6 and 6.2 keV 
at $> 99$\% confidence. 
The ``difference spectrum'' (Figure 2d)   i.e.,  
the spectrum from T2 minus that from T1,  further highlights the 
variable components, confirming changes in the line profile at $> 99$\% 
confidence. 
The apparent variability of the 5.6 keV line may be due to 
a shift of the line peak from 5.4 to 5.6 keV (the 
{\it HETG} data is weighted to a later time). 
Binning the line coarsely illustrates that  {\it Chandra} (Figure 2e)
and {\it XMM} (Figure 2f) data reveal a broad component (which will be 
detailed in a later paper) 
of equivalent width 
$\sim 800$ eV (using a Gaussian model, addition of which 
yielded $\Delta \chi^2=200 $ for $1483\ dof$).  
Figure 2f shows the broad component of the 
line maintains a $\sim$constant equivalent 
width, while that of the narrow line changes from 216 eV (T1) to 172 eV (T2) 
as the continuum changes flux. 

\section{Discussion} 

We now consider the origin of the newly discovered features. 
While 6.2 keV is the energy of the first peak of a Compton shoulder, the 
observed line is too sharp and strong to be attributed to this. 
It has been 
proposed \citep{skibo97} that the destruction (spallation) of Fe into Cr 
and other lower Z metals on 
the surface of the accretion disk enhances  the line emission 
expected from elements of 
low abundance. In this way lines at $\sim 5.6$ and 6.2 keV  
from Cr{\sc xxiii} and Mn{\sc xxiv}  
could be enhanced to a detectable level. The observed Cr to Mn  
ratio, however, is 
inconsistent with the spallation model \citep{skibo97}. 
%which considers a wide  range of physical conditions. 
The lines are most likely due to Fe, shifted by 
relativistic effects. 

Absorption from an infalling blob of gas was previously 
suggested to explain the complex and variable 
profile observed with {\it ASCA} \citep{n99}. 
However, comparison of Figs 2b and 2c demonstrates that the complexity of the 
profile is due to the appearance of emission features during some time intervals. 
If the emission line at 5.6 keV is from infalling neutral gas, the implied velocity
must be $ > 40,000\ {\rm km\ s^{-1}}$.   
Infalling gas would likely 
have to be ionized such that resonance line and bound-free opacity 
were small enough that 
the gravitational attraction of the disk/black hole
system could overcome radiation pressure \citep{psk00}, thus flow 
velocities would be even higher. 
However, the apparent shifts in the line
energy (5.4 -- 5.6 keV) with time are more indicative of deceleration of an
outflowing blob. 

Recently, {\it Chandra} observations of the Galactic X-ray binary SS~433
revealed relativistically red- and blue-shifted lines from 
Fe{\sc xxvi} Ly-$\alpha$ (6.97 keV) and Fe{\sc xxv} 
1s2p-1s$^{2}$ (6.69 keV), with an outflow velocity of 
$\approx$ 0.27$c$ \citep{marshallea02}.
The narrow lines at 5.6 eV and 6.2 keV cannot be 
explained by H- and He-like Fe emission from a single flow.
However, if the lines are Ni{\sc xxvii} 1s2p-1s$^{2}$ (7.789 keV) and 
Fe{\sc xxvi} Ly$\alpha$, the redshifts are 
consistent with a velocity $\sim 0.25$c (assuming a disk inclination of
38$^{\rm o}$ \citealt{wh}). In fact, 
\citet{marshallea02} 
report the detection of relativistically red-shifted Ni{\sc xxvii} in 
SS~433. 
\citet{wangea00} predict that blobs of gas originating from disk
instabilities 
will be initially be fully ionized, but will show line emission via
recombination  sometime after ejection ($\sim$ days, assuming
T $\sim 10^{8}$K, and n$_{e}$ $\sim$ 10$^{10}$ cm$^{-3}$). 
Hence it is possible that these emission lines are recombination lines of
He-like Ni and H-like Fe, although we require an overabundance of 
Ni or a peculiar
ionization balance in the recombining gas. Since there are no 
blue-shifted features (at 8.21, 9.19 keV keV) 
the jet must be one-sided, perhaps an example of
the MHD driven jets suggested by \citet{cbt96}. 
As we see the red side, our line-of-sight to the 
jet, presumably through the inner regions of the accretion disk, cannot be
blocked by a Compton-thick layer. 

It has been debated \citep{missut,f95,rw2000,rszk,mis01} 
whether  thermal Comptonization of a narrow Fe K$\alpha$ line could explain 
the broad components in AGN.  The presence of sharp line features means no  
Comptonizing medium exists between us and the nuclear environs. Thus the
broad line 
in NGC 3516, and probably other AGN, must be produced by general
relativistic effects.  The alternative suggestion, that 
the broad feature is an 
absorbed continuum component, does not allow us to avoid 
invoking relativistic effects to explain the sharp features. 

The constraints on the velocity widths of the new features indicate they originate 
in material with an organized 
kinematic  structure such as the accretion disk, 
arising from narrow annuli or hotspots 
\citep{sem99,nk01}.  
Reverberation of flares across the disk predicts narrow features \citep{young2000} 
but they should only exist for a few hundred 
seconds.  If the gas is ionized, the 
line could be emitted at 7 keV  
requiring extreme relativistic shifts in the maximally-rotating Kerr 
metric \citep{L91}, viewed at 
low inclinations 
to match observed features. Assuming the central regions of
NGC~3516 to be inclined at 
38$^{\rm o}$ \citep{wh}, 
the lines would be produced between 3 - 5 gravitational radii ($R_g$) 
where relativistic effects predict lines 
significantly broader and less ``peaky'' than observed, assuming we 
see a full orbit of the material round 
the black hole. 
 Orbital timescales ($t_{\rm orb} \sim ({\rm r/9 R_g})^{3/2} {\rm M_8} $) are 
 1 -- 2.3 hours for 3 -- 5 ${\rm R_g}$ (around black hole mass 
 $M \sim  2.3 \times 10^7 M_{\odot}$ estimated for NGC~3516;
\citealt{dinella}). Our spectra sample several orbits of material this close 
to the black hole.  Thus 
line widths allow us to rule out an origin in an 
ionized disk around a spinning black hole. 

Alternatively, the features arise in cool gas, where 
the required energy shifts are smaller. 
The peak at 5.6 
keV could be the red horn of a line from $\sim 35\ R_g$   
(a weak  peak close to 6.8  keV may be 
the blue horn associated with this). The 6.2 and 6.5 keV  
peaks could be due to emission from $\sim  175\ R_g$ (Figure~4). 
At these radii the 
feature widths match predictions for illuminated annuli 
or orbiting blobs ($175\ R_g$ may be outside the disk structure).   
Emission from a restricted area on the disk 
was previously suggested by \citet{i99} based on a flare state in  
MCG-6-30-15. Thus the phenomenon observed here 
may be generally applicable to AGN. 
%Precise determination of the putative blue horns is difficult, our 
%data are qualitatively 
%consistent with the proposed origin in a cool disk. 
The lack of strong features blueward of 6.4 keV is a problem for 
all the models discussed here. 
Occultation (\citealt{my98})
may be important, where different regions of 
the disk become visible at different times.

In summary, new data show spectral features most likely 
explained as Fe K$\alpha$ lines, modified by relativistic effects. 
An increase in the X-ray 
continuum flux is correlated with the detection of these lines.  Regions of 
magnetic  reconnection  would illuminate very small areas of the accretion 
flow and may 
provide the source of enhanced continuum emission. 
Reconsideration of the {\it ASCA} data indicates that 
the phenomenon driving the variations in 
line profile has been going on for at least 
several years and over a range of nuclear luminosity.

\section{Acknowledgements}

We are grateful to the {\it Chandra}, {\it XMM} and {\it RXTE}  
satellite operation teams for co-ordination 
of the multi-waveband observations. 
We thank Craig Markwardt for production of 
the {\it RXTE} spectrum. 
T.J.\ Turner acknowledges support from NASA 
grant  NAG5-7538. 
This research was supported by the Israel Science Foundation (grant
no. 545/00). 

\clearpage

\begin{deluxetable}{lccc}	%%%%%%%%%%%%%%%%%%%%%%%%%%%%%%%%%%%%%%%%%%%
\tablewidth{0pc} 
\tablecaption{HETG Narrow Components of Fe K$\alpha$}
\tablehead{\colhead{Energy (keV)} & \colhead{Flux$_{\rm T1-XMM}$} &
	 \colhead{Flux$_{\rm T2-XMM}$}      & \colhead{Flux$_{\rm HETG}$} \\
}    
\startdata

$5.57\pm0.02$ &		 $0.00(< 0.53)$ & 		$0.00 (< 2.89)$ &  
	$7.37 (5.90 - 8.60)$ \nl 
$6.22\pm0.04$ &		$6.26 (5.07-7.94)$ &	$9.90 (5.92-13.3)$ & 
	$11.1 (9.45 - 13.0)$ \nl
$6.41\pm0.01$ &		$36.5 (34.8-38.5)$ &	$33.5(28.7-37.7)$ &
	$31.7 (26.9  - 34.1)$ \nl
$6.53\pm0.04$ &	       $8.19 (6.94 - 10.2)$ &	$8.98(4.67 - 12.6)$ &	
	$7.43 (5.47 - 9.95)$ \nl
$6.84-6.97$ &	  $5.12 (2.75 - 5.99)$	& $5.78(0.97 - 8.18)$ &
	 \nodata \nl 

\tablenotetext{a}{Errors are $\chi^2 + 2.71$ with line energies 
fixed and narrow (except for 6.4 keV where the fitted width
was fixed). Line fluxes in units 
$\times 10^{-6}$ photons cm$^{-2}$ s$^{-1}$. 
We tabulate fluxes for lines appearing at $6.84\pm0.01$ (T1) and $6.97\pm0.06$ keV 
(T2) in {\it XMM}  
data.  A simple broad Gaussian 
component was included to parameterize the broad line, with 
$E=5.19\pm0.17$ keV, $\sigma=1.19\pm0.19$ eV, 
flux n=$1.81^{+0.37}_{-0.29} \times 10^{-4}$ photons cm$^{-2}$ s$^{-1}$. }
\enddata 
\end{deluxetable}

\clearpage

\typeout{FIGS}

\begin{figure}	 %%%%%%%%%%%%%%%%%%%%%%%%%%%%%%%%%%%%%%%%%%%
	\epsscale{0.9} 
\includegraphics[scale=0.70,angle=0]{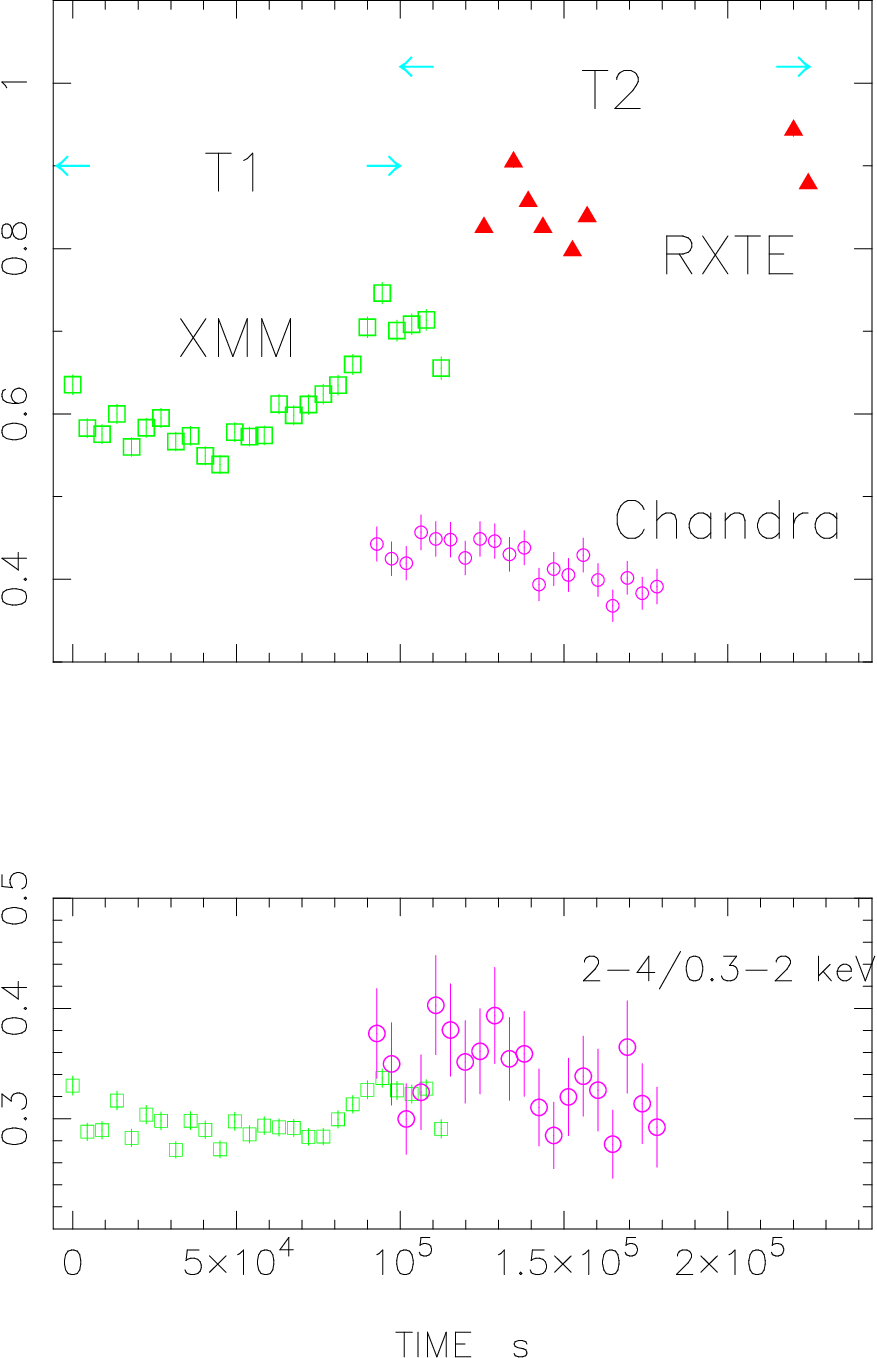}
%	\plotone{lc_fig.cps}
	\caption[Light Curves]
{The upper panel shows light curves from {\it XMM} (2-6 keV, 
green squares);  
{\it Chandra} (2-6 keV, magenta circles) and {\it RXTE} (3-30 keV, red
triangles). 
The time is shown in seconds relative to the start of the 
{\it XMM} observation at 2001 November 09 UT 23:12:51. The 
{\it Chandra} data have been scaled up by a factor 4, the 
{\it RXTE} data have been scaled down by a factor 12.5, for 
clarity of display. 
The lower panel shows the hardness ratio, 2-4/0.3-2 keV, 
based on the {\it XMM} PN data and {\it Chandra} ACIS data.
The series are in 4500 s time bins and the {\it Chandra} ratio has been
scaled down by a factor of 4, for display. 
\label{fig:lc} 
}
\end{figure}

\begin{figure}
\includegraphics[scale=0.35,angle=0]{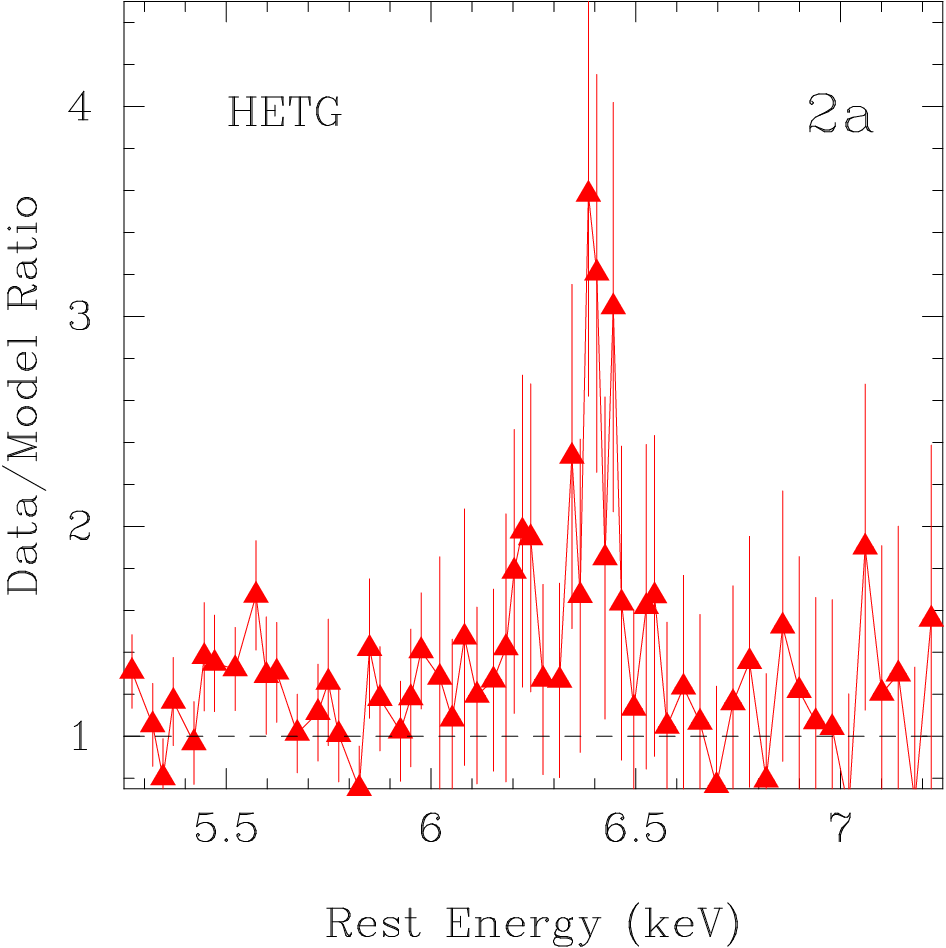}
\includegraphics[scale=0.35,angle=0]{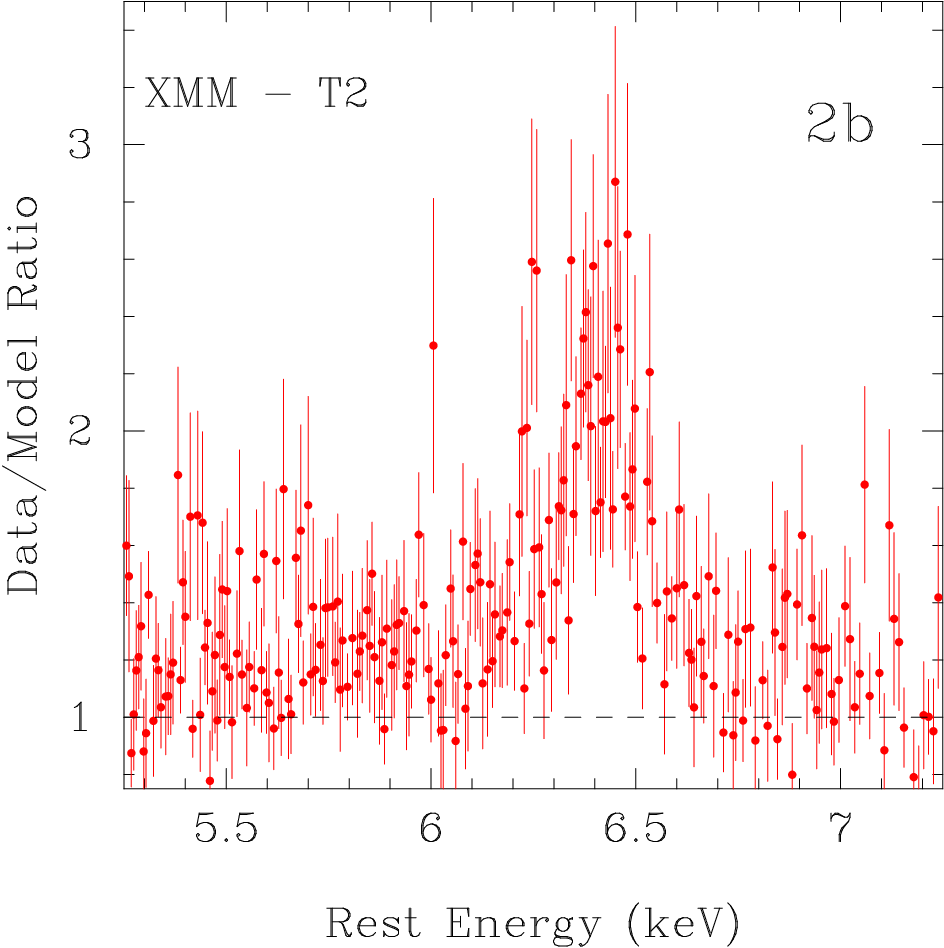}
\includegraphics[scale=0.35,angle=0]{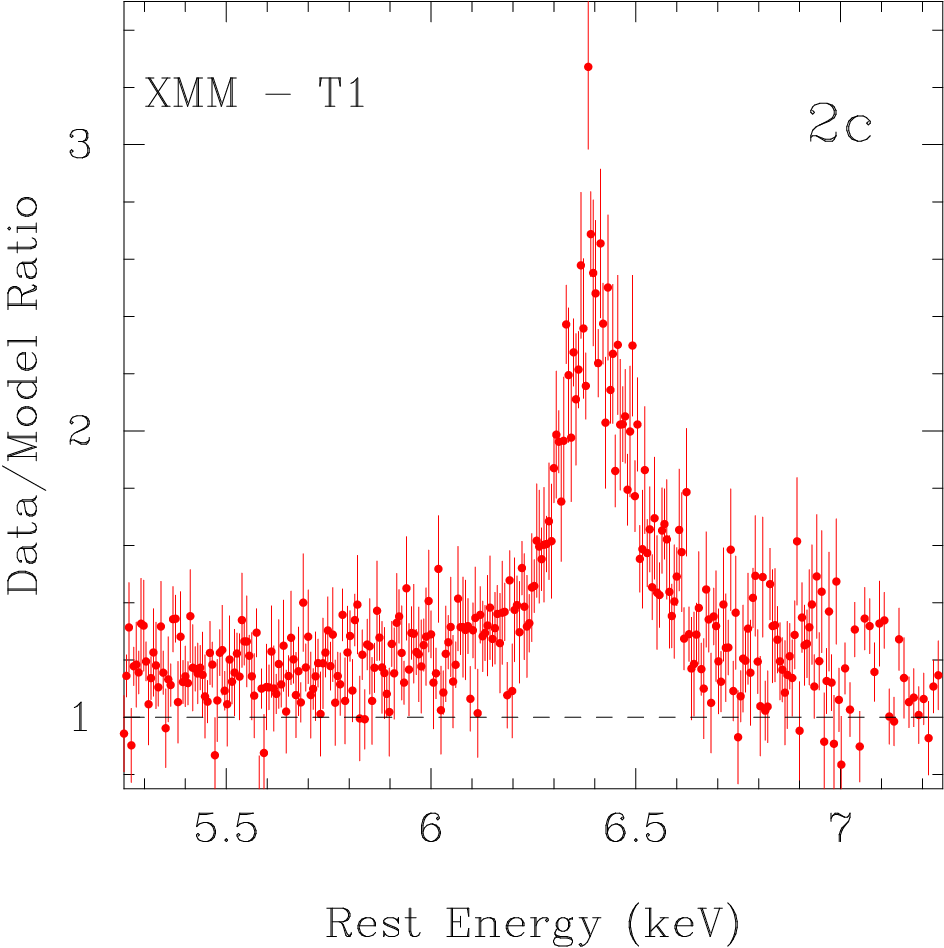}
\includegraphics[scale=0.35,angle=0]{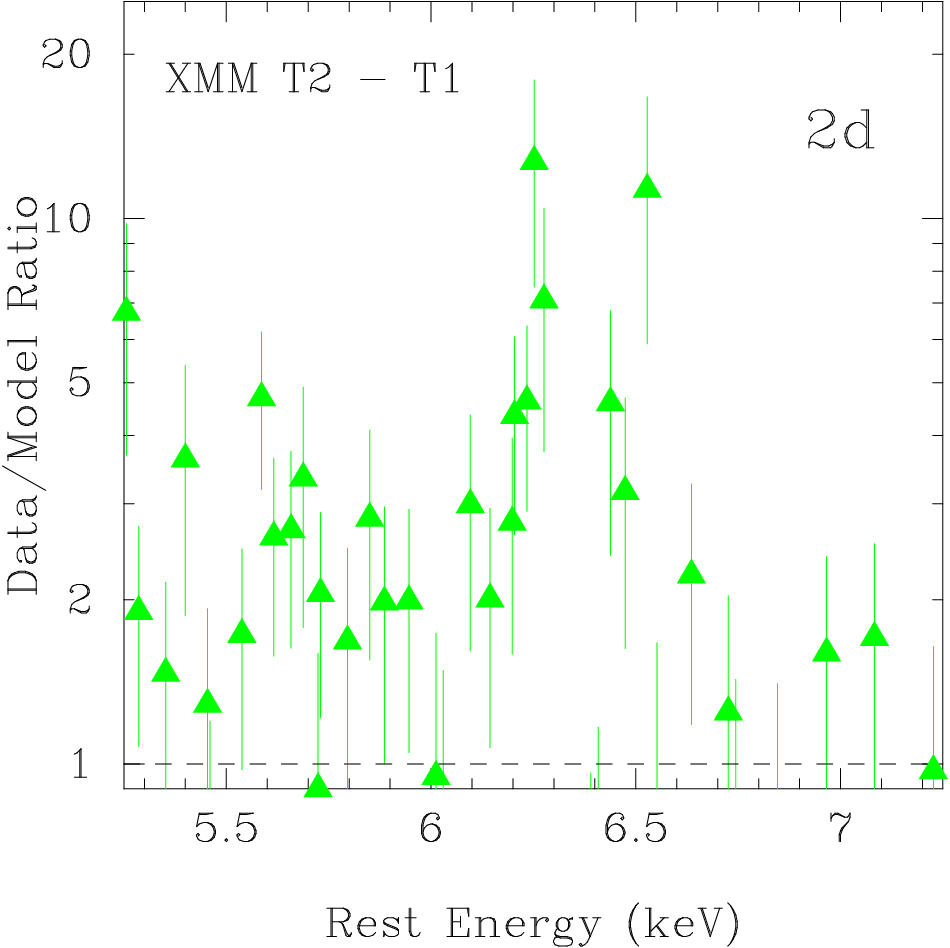}
\includegraphics[scale=0.35,angle=0]{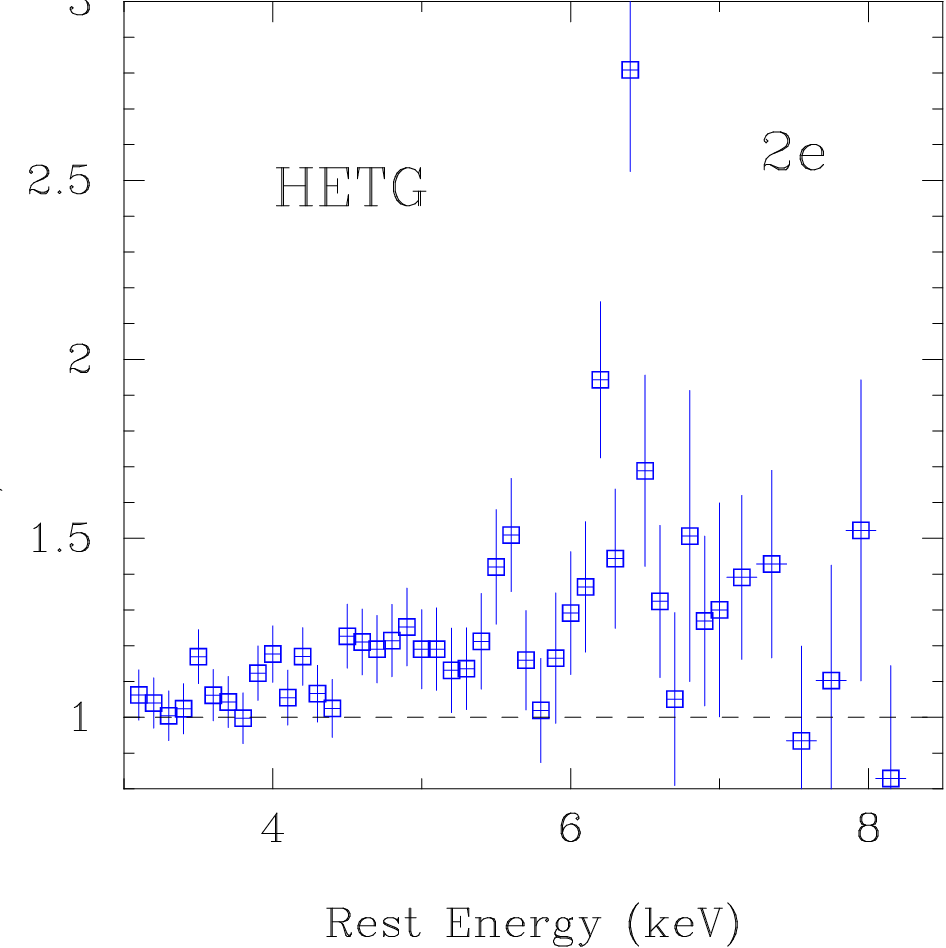}
\includegraphics[scale=0.35,angle=0]{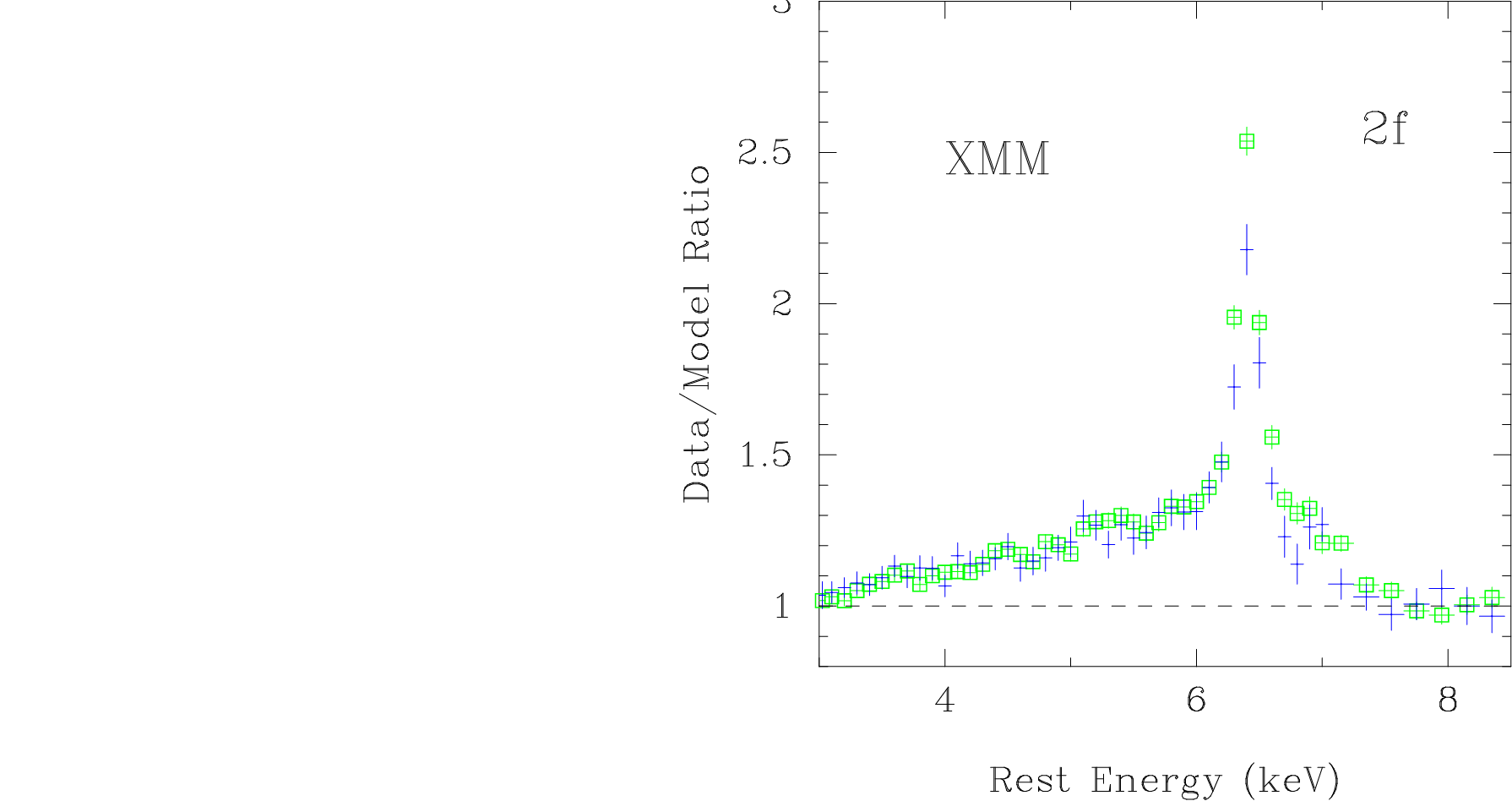}
\caption{
The X-ray data/model ratios in the Fe K$\alpha$ regime. 
The model is a power-law continuum with $\Gamma \sim 1.2$. 
Data from the high-state with 
{\it HETG} (a) and {\it XMM} (b). HETG data 
are the sum of the positive and negative 1st order 
grating spectra plus MEG data up to 6.5 keV. 
Also {\it XMM} data from the low-state (c) 
and the difference spectrum (the high-state minus the 
renormalized low-state spectra for PN, MOS1 and MOS2) compared 
to a power-law fit (d; {\it XMM} data alone). 
e) shows the coarsely-binned 
HETG data; (f) the {\it XMM} data 
(T2 - blue squares, T1 - green crosses); 
we combine data from PN, MOS1 and MOS2, which 
agree when viewed individually. }

\label{fig:fek}
\end{figure}
\clearpage

\begin{figure}	 %%%%%%%%%%%%%%%%%%%%%%%%%%%%%%%%%%%%%%%%%%%
\includegraphics[scale=0.50,angle=0]{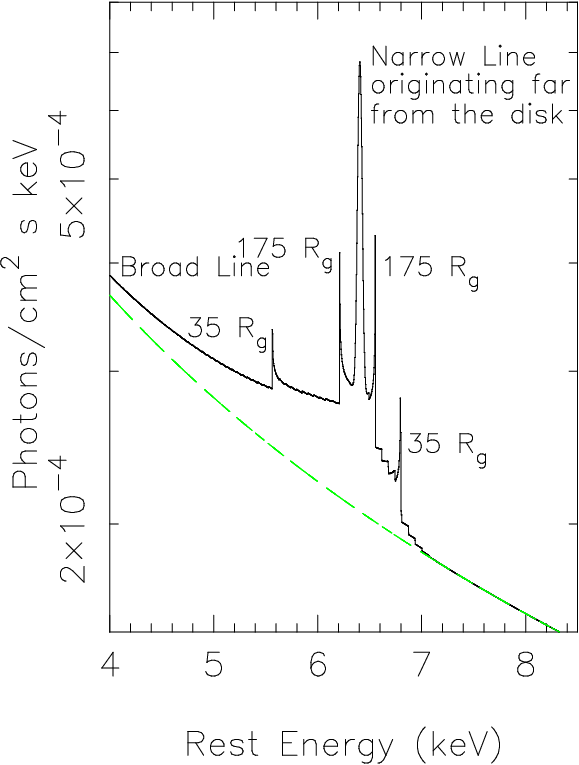}
	\caption[Schematic Fe Ka]
{A schematic representation of the components of the Fe K$\alpha$ 
line in NGC~3516. The green line shows the continuum level. 
\label{fig:schematic_dl}
}
\end{figure}

\end{document}